\documentclass[%
  preprint,
  superscriptaddress,
  amsmath,amssymb,
  aps,
  pre,
]{revtex4-1}

\usepackage{bm}

\usepackage{graphicx}
\usepackage{dcolumn}
\usepackage{bm,ulem,color}
\usepackage{epstopdf}
\usepackage[utf8]{inputenc}
\usepackage{epstopdf}

\usepackage{color}

\usepackage{multirow}

\begin{document}

\title{Inverse energy cascade in nonlocal helical shellmodels of turbulence \footnote{Version accepted for publication (postprint) on  Phys. Rev. E 92, 043021 (2015) – Published 28 October 2015}}

\author{Massimo De Pietro} 
\affiliation{Dip. di Fisica and INFN,
  Universit\`a ``Tor Vergata", Via della Ricerca Scientifica 1,
  I-00133 Roma, Italy.}  

\author{Luca Biferale} \affiliation{Dip. di
  Fisica and INFN, Universit\`a ``Tor Vergata", Via della Ricerca
  Scientifica 1, I-00133 Roma, Italy.}  
  
\author{Alexei A. Mailybaev}
\affiliation{Instituto Nacional de Matem\'{a}tica Pura e Aplicada-IMPA, Est. Dona Castorina 110, Rio de Janeiro 22460-320 Brazil}

\date{\today}

\begin{abstract}
Following the exact decomposition in eigenstates of helicity for the Navier-Stokes equations in Fourier space  [F. Waleffe, \textit{Phys. Fluids A} \textbf{4}, 350 (1992)] we introduce
a modified version of helical shell models for turbulence with non-local triadic interactions. 
By using both an analytical argument and numerical simulation, we show that there exists a class of models, with a specific helical structure, that exhibits a statistically stable inverse energy cascade, in close analogy with that predicted for the Navier-Stokes equations restricted to the same helical interactions.  We further support the idea that  turbulent energy transfer is the result of a strong entanglement among triads  possessing different transfer properties. 
\end{abstract}

\maketitle

\section{Introduction}

Understanding and controlling the statistical and dynamical properties of turbulent flows is still an open problem in many fundamental and applied fields. 
From a theoretical point of view, the main difficulties stem from the highly non-linear nature of the dynamics
in the fully developed regime. Moreover, the  presence of a large separation between the injection and dissipative scales and  the empirical observation of non-Gaussian statistics of the velocity field make the system hard to approach with analytical perturbative techniques 
or brute force direct numerical simulations \cite{pope_turbulent_flows,frish_turbulence}.
The physics of a turbulent flow is very rich. It might depend on the embedding dimensionality, leading to a direct transfer of energy from large to small scales in three dimensions (forward cascade) or to an inverse transfer in two dimensions (backward cascade). Moreover, in the direct regime, turbulent flow develops  anomalous scaling laws, where different moments of the velocity fluctuations possess a power-law behavior as a function of the separation scale, characterized by a set of anomalous scaling exponents. 

For these reasons many different techniques and approximations have been developed in order to try to better understand the turbulent phenomenology.
 One such approach is represented by shell models \cite{obukhov1971some, gledzer1973system, desnianskii1974evolution, yamada1988inertial, jensen1991intermittency, Lvov_1998_improved_shellmodels, biferale2003shell, bohr2005dynamical, ditlevsen2010turbulence}.

Shell models of turbulence are simplified models that mimic the Navier-Stokes (NS) equations in wave-number space. They are based on a  strong reduction in the number of degrees of freedom, keeping 
only a few  representative variables  (typically one or two real variables) for the whole original set of wave-numbers belonging to each shell.  To have scaling invariance embedded in the system, 
the shell variables are defined on a set of wavenumbers equally spaced on a logarithmic 
scale, $k_n \sim \lambda^n k_0$, where $\lambda=2$ conventionally. In this way, a large separation of scales is achieved with relatively few variables. 
Furthermore, inspired by Kolmogorov phenomenology for the direct energy transfer, these models
 consider only local interactions in Fourier space, connecting dynamical evolution between three neighboring modes  $k_n, k_{n+1}, k_{n+2}$. Last but not least, the models are built in such a way that they have the same inviscid invariants of the original Navier-Stokes equations: energy and helicity for  models of three-dimensional(3D) turbulence or energy and enstrophy for the 2D case. 

Despite the huge simplifications, shell models share many properties with the original Navier-Stokes turbulence, including the development of anomalous scaling laws with values of the scaling exponents very close to the ones measured in  3D turbulence \cite{jensen1991intermittency, pisarenko1993further, Lvov_1998_improved_shellmodels, biferale2003shell}. Many generalizations to models for magnetohydrodynamics \cite{plunian2013shell}, rotating fluids \cite{reshetnyak2003shell, hattori2004shell, chakraborty2010two}, convection \cite{brandenburg1992energy, Mingshun1997Scaling, Ching2008Anomalous, Ching2008Refined, Kumar2015Shell} and passive scalars \cite{jensen1992shell, benzi1997analytic, arad2001statistical} have also been studied. 

Notwithstanding their success, shell models proved to be problematic when inverse energy cascade becomes the dominant phenomenon to be studied. 
In fact, in all known  models for 2D turbulent flows that conserve energy and 
 enstrophy  the inverse energy flux 
 is overwhelmed by  equilibrium fluctuations \cite{aurell1994statistical, gilbert2002inverse}. 
Similarly, also considering shell models of 3D Navier-Stokes equations restricted to having only sign-definite helicity \cite{biferale2012inverse}, the inverse energy cascade is sub-leading with 
respect to equilibrium fluctuations \cite{gilbert2002inverse}. 
Indeed, an inverse energy cascade in shell models has been observed 
only by adding  extra terms in the equations of motion, representing mechanisms such as rotation or stratification \cite{chakraborty2010two, Kumar2015Shell}, or considering the dynamics 
in a range of parameters where the conserved quantities have different physical dimensions with respect to those of the Navier-Stokes equations \cite{gilbert2002inverse}. The main goal of this paper is to present a  shell model that has energy and helicity as inviscid invariants, and that shows an inverse energy cascade without relying on any additional external  mechanism beside the ones already present in the NS non-linear term.

To better understand the interplay between helicity and energy, shell models for 3D turbulence have been proposed in \cite{benzi_1996_Helical_shell_models} using a close connection with the helical structure of the original Navier-Stokes equations. The idea was to apply
 the decomposition in helical eigenstates of the Navier-Stokes equations in order to distinguish triadic non-linear interaction on the basis of
 their helical content \cite{waleffe_1992_The_nature_of_triad_interactions}. 
It was indeed argued in \cite{waleffe_1992_The_nature_of_triad_interactions} that depending on the relative sign of helicity carried by the three interacting modes, energy tends to be transferred forward or backward in 3D turbulent flow. Recently, further support for this statement was given in \cite{biferale2012inverse} by performing direct numerical simulations of 3D turbulence under the constraint of having only sign-definite helical modes and showing that in this case the flow inverts the energy transfer direction, by pumping energy to larger and larger scales. As a result,  clear evidence that inverse and direct energy transfer mechanisms might coexists in 3D turbulence was given, making it even more interesting to understand under which circumstances the former prevails over the latter, or vice versa. 

In this paper, we expand the work done in \cite{benzi_1996_Helical_shell_models}, trying to understand if the inclusion of helical variables in shell models might shed some light on the complexity of the energy transfer mechanism. In particular, we show that the \textit{aspect ratio} of the triads is a key point.  To achieve an inverse energy transfer mechanism we relaxed the constraint of first-neighbor interactions between wave-numbers. Indeed,  we show, with both theoretical and numerical tools, that this simple modification can have dramatic consequences on the energy-cascade mechanism, turning a model that exhibits direct energy cascade into a model that exhibits an inverse energy cascade.
It is remarkable that the argument suggesting the importance of elongated triads is taken in full similarity with the original case of 3D Navier-Stokes equations as developed originally in  \cite{waleffe_1992_The_nature_of_triad_interactions}:  another case of a  close overlap between the physics of turbulence and the dynamics of shell models. 

The paper is organized as follows. In Sec. \ref{sec:model_definition}, the helical decomposition is briefly reviewed and a modified SABRA model with more \textit{elongated} triads is defined. In Sec. \ref{sec:energy_transfers}, predictions for the direction of the energy cascade and scaling laws are made, on the basis of the stability analysis of a single interacting triad. In Sec. \ref{sec:simulations_single}, results of numerical simulations are shown and compared with the predictions from the previous section. Finally, the two appendixes contain details and calculations. Appendix \ref{sec:app_general_sabra_equations} contains the definition of a more general helical shell-model, with triads of any shape. Appendix \ref{sec:app_instability_assumption} contains detailed calculations for the stability analysis of a single interacting triad.

\section{Helical decomposition for shell models of turbulence}
\label{sec:model_definition}

\subsection{The original SABRA model}
The original SABRA shell model \cite{Lvov_1998_improved_shellmodels} was inspired by the Navier-Stokes equations in Fourier space, and, although it cannot be formally derived from them, it has a phenomenology very similar to that of 3D homogeneous and isotropic turbulent flows. The model 
describes the evolution of a single complex variable $u_n$, representing all the modes in a shell of wave-numbers $ |\mathbf{k}| \in [k_n,k_{n+1}]$. The equations of motion take the form \cite{Lvov_1998_improved_shellmodels}:
\begin{equation}
\label{eq:sabra_standard}
\dot{u}_n = \, i (a k_{n+1} u_{n+2} u_{n+1}^{*} + b k_{n} u_{n+1} u_{n-1}^{*} + c k_{n-1} u_{n-1} u_{n-2}) - \nu k_n^\beta u_n + f_n +\nu_l k_n^{-4} u_n\, ,
\end{equation}
where, $k_n=k_0 \lambda^n$, $\lambda$ is an arbitrary scale parameter larger than unity (here $\lambda=2$), $ \nu k_n^{\beta} $ is a dissipative ($\beta=2$) or hyper-dissipative ($\beta >2$) term, $f_n$ is an external forcing term, and $\nu_l k_n^{-4} $ is a large-scale damping term introduced  for those models
that develop an inverse energy transfer in order to get a stationary statistics. The model is defined on a given number of shells, $n=0,1,\dots, N$, and the 
boundary conditions $u_{-1}=u_{-2}= u_{N+1}=u_{N+2}=0$ are imposed. 
 
 The model has two quadratic inviscid invariants that depend on the values of the  $a,b,c$ parameters. The first one is always chosen to be the energy, $E = 
\sum_{n=0}^N |u_n|^2$, while the second can be defined to be unsigned to mimic helicity in 3D Navier-Stokes 
equations, $H = \sum_{n=0}^N (-)^n k_n |u_n|^2$, or 
positive definite as enstrophy for 2D NS, $\Omega = \sum_{n=0}^N k_n^2 |u_n|^2$. A significant drawback of the above model in the 3D regime is the imbalance between successive shell variables: the ones with an odd shell index carry only negative helical modes, while the ones with even $n$ carry positive helicity 
\cite{ditlevsen2001cascades,chen2003joint, lessinnes2011dissipation}.

\subsection{The helical SABRA model}
\begin{figure}[hbt]
\centering
\includegraphics{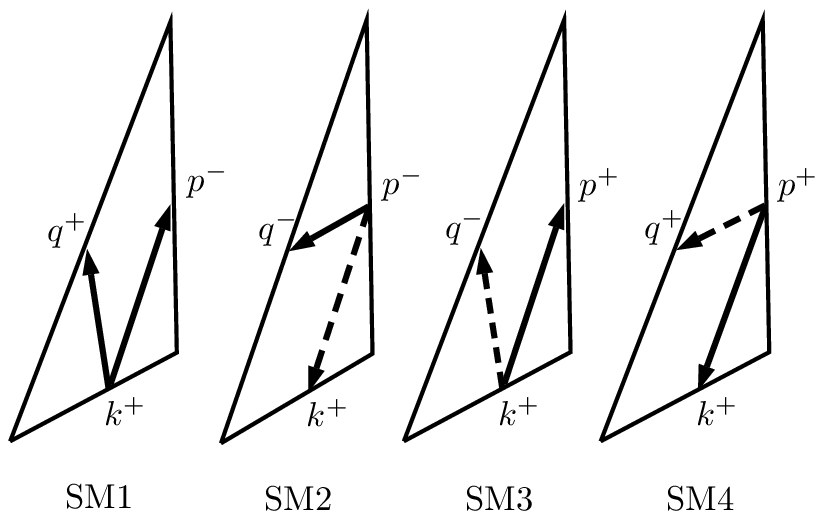}
\caption{Representation of the four independent classes of helical interaction between an ordered triad of wave-numbers $k<p<q$, in both Navier-Stokes and helical shell models. The $\pm$ superscripts represents the helical mode which is participating in the interaction. Each class has two possible interactions, that are equivalent due to the parity symmetry $k^+ \rightarrow k^-, p^+ \rightarrow p^-, q^+ \rightarrow q^-$; only one is shown here. The arrows represent the energy transfers, as a result of the instability assumption (see section \ref{sec:energy_transfers} and Appendix \ref{sec:app_instability_assumption}). The dashed arrows represent weaker transfers with respect to the full lines. For models 1 and 3 energy flows out of the smallest wave number, in particular in model 1 the smallest wave-number transfers the same amount of energy to the other two, while model 3 exhibits a more localized energy transfer; in model 2 the middle wave-number transfers more energy to the largest wave-number and less to the smallest; in model 4 the middle wave-number transfers more energy to the smallest wave-number and less to the largest.}
\label{fig:triad_stability}
\end{figure}

 In order to overcome the previous limitation a new class of shell models with a more 
realistic helicity structure was proposed in \cite{benzi_1996_Helical_shell_models}. The first step was to follow the exact decomposition of the Navier-Stokes velocity field, in Fourier space, into positive and negative polarized helical waves \cite{waleffe_1992_The_nature_of_triad_interactions}:
\begin{equation}
\label{eq:waleffe_ns_u_decomposition}
\mathbf{u}(\mathbf{x}) = \sum_{\mathbf{k}} (u_{\mathbf{k}}^+ \mathbf{h}_{\mathbf{k}}^+ + u_{\mathbf{k}}^- \mathbf{h}_{\mathbf{k}}^-) e^{i \mathbf{k} \cdot \mathbf{x}} \, ,
\end{equation}
where $\mathbf{k}, \mathbf{h}_{\mathbf{k}}^+, \mathbf{h}_{\mathbf{k}}^-$ form an orthogonal basis, and the two $\mathbf{h}_{\mathbf{k}}^s$ (with $s=\pm$) are eigenvectors of the curl operator:
\begin{equation}
i \mathbf{k} \times \mathbf{h}_{\mathbf{k}}^s = s k \mathbf{h}_{\mathbf{k}}^s \, .
\end{equation}
A possible way to construct them is to use the decomposition: 
\begin{equation}
\label{eq:waleffe_h_s}
\mathbf{h}_{\mathbf{k}}^s = \boldsymbol{\nu}_{\mathbf{k}} \times \boldsymbol{\kappa} + i s \boldsymbol{\nu}_{\mathbf{k}} \, ,
\end{equation}
where $\mathbf{k} = k \boldsymbol{\kappa}$, $\boldsymbol{\nu}_{\mathbf{k}} = (\mathbf{z} \times \boldsymbol{\kappa}) / ||\mathbf{z} \times \boldsymbol{\kappa}||$, and $\mathbf{z}$ is an arbitrary vector. The two fields $u^+_{\mathbf{k}}$ and $u^-_{\mathbf{k}}$ are nothing more than the projections on the $\mathbf{h}_{\mathbf{k}}^+$ and $\mathbf{h}_{\mathbf{k}}^-$ directions of the Fourier coefficients of the velocity field, and they carry, respectively, positive and negative helicity.
It was realized that by plugging this decomposition into the non-linear term of the Navier-Stokes equations, one can distinguish eight possible non-linear triadic interactions depending on the signs of the corresponding helical 
projections \cite{waleffe_1992_The_nature_of_triad_interactions}. Four out of eight interactions are independent, because the interactions with reversed helicities are identical; they are summarized in Fig. \ref{fig:triad_stability}. It is possible to apply the same decomposition \textit{verbatim} to construct different classes of helical shell models with a more accurate helical structure than the original model \eqref{eq:sabra_standard}. This first step was done in \cite{benzi_1996_Helical_shell_models} introducing two complex variables $u_n^+$ and $u_n^-$ for every wave-number, each one of them carrying positive or negative helicity and leading to the four independent classes of the local helical shell-model. All of them have the form:
\begin{align}
\label{eq:sabra_helical_standard_up}
\dot{u}_n^+ = & \, i (a k_{n+1} u_{n+2}^{s_1} u_{n+1}^{s_2*} + b k_{n} u_{n+1}^{s_3} u_{n-1}^{s_4*} + c k_{n-1} u_{n-1}^{s_5} u_{n-2}^{s_6}) - \nu k_n^\beta  u_n^+ + f_n^+ -\nu_l k_n^{-4} u_n^+ \, ,\\
\label{eq:sabra_helical_standard_um}
\dot{u}_n^- = & \, i (a k_{n+1} u_{n+2}^{-s_1} u_{n+1}^{-s_2*} + b k_{n} u_{n+1}^{-s_3} u_{n-1}^{-s_4*} + c k_{n-1} u_{n-1}^{-s_5} u_{n-2}^{-s_6}) - \nu k_n^\beta   u_n^- + f_n^-   -\nu_l k_n^{-4} u_n^- \, ,
\end{align}
where the  helical indices $s_i = \pm$ are reported in Table \ref{tab:helicity_indices} and the coefficients $a,b,c$ can be found in Table \ref{tab:model_coefficients}. Each one of these models evolves according to only one of the four independent helical interactions depicted in Fig. \ref{fig:triad_stability}, where a triad $(k_{n-1},k_n,k_{n+1})$ is represented by $(k,p,q)$.

It is important to stress that, exactly as in the original Navier-Stokes equations, the four classes of interactions conserve energy and helicity separately, if the coefficients $a,b,c$ are chosen appropriately, i.e. they can  be considered as four sub-models of the whole problem. The added value with respect to the previous SABRA structure is that now energy and helicity have the very same structure as for the NS equations \cite{waleffe_1992_The_nature_of_triad_interactions}
without the asymmetry among odd and even shells: 

\begin{eqnarray}
\label{eq:energy_def}
E = \sum_{n=0}^N ( |u_n^+|^2 + |u_n^-|^2) \, ,\\
\label{eq:helicity_def}
H = \sum_{n=0}^N k_n( |u_n^+|^2 - |u_n^-|^2).
\end{eqnarray}

As we shall see later, none of these 4 models is indeed able to show an inverse energy cascade. Even the shell model SM4, which is the equivalent of the Navier-Stokes restriction to sign-definite helical interactions \cite{biferale2012inverse}, fails to develop an inverse energy transfer because of the presence of strong fluctuations due to the quasi-equilibrium solution \cite{gilbert2002inverse}. It is not surprising that the equilibrium solution might have a different influence on the shell model with respect to the Navier-Stokes equations, because of the strong difference in the scaling of the number of degrees of freedom as a function of the
embedding physical dimension. A priori there is no reason why a very simplified structure such as the one given by shell models should replicate exactly the behavior of the Navier-Stokes equations restricted on the same helicity interactions class. In particular, one of the strongest limitations is given by the restriction to very local interactions among Fourier variables assumed by the structure \eqref{eq:sabra_helical_standard_up}. 

In \cite{waleffe_1992_The_nature_of_triad_interactions} it was shown, on the basis of an ``instability assumption'', that triads where the two highest wave-numbers have the same helical sign, such as those in model SM2 (see Fig. \ref{fig:triad_stability}), might lead to an inverse cascade. It was also explained that the key factor for the NS case is the triad  geometry. Calling $v=k/p$ the ratio between the smallest and middle wave-number, it was argued in \cite{waleffe_1992_The_nature_of_triad_interactions}, on the basis of a phenomenological scaling argument, that if $v < 0.278$, the triad should contribute to an inverse flux of energy, from small to large scales.
In fact, empirical observation made on direct numerical simulations of the shell-model SM2 (where $v=0.5$) showed that energy flows toward small scales. 
We are therefore interested in extending the range of interactions, exploring smaller values of the ratio $v$, in order to meet  the ``elongation'' requirement argued in \cite{waleffe_1992_The_nature_of_triad_interactions}.

\begin {table}[Hb]
\caption{Helicity indices of Eqs. \eqref{eq:sabra_helical_standard_up}-\eqref{eq:sabra_helical_standard_um} and \eqref{eq:sabra_helical_general_up}-\eqref{eq:sabra_helical_general_um} for the four models.}
\label{tab:helicity_indices} 
\begin{tabular*}{\linewidth}{@{\extracolsep{\fill} } c  c  c  c  c  c  c }
    \toprule
    model & $s_1$ & $s_2$ & $s_3$ & $s_4$ & $s_5$ & $s_6$ \\ \colrule
   SM1   &  $+$ &  $-$ &  $-$ &  $-$ &  $-$ &  $+$ \\
   SM2-SM2E   &  $-$ &  $-$ &  $+$ &  $-$ &  $+$ &  $-$ \\
   SM3   &  $-$ &  $+$ &  $-$ &  $+$ &  $-$ &  $-$ \\
   SM4   &  $+$ &  $+$ &  $+$ &  $+$ &  $+$ &  $+$ \\
    \botrule
\end{tabular*}
\end{table}

\begin {table}[Hb]
\caption{Coefficients of equations \eqref{eq:sabra_helical_standard_up}-\eqref{eq:sabra_helical_standard_um} for the four helical shell models with first-neighbor interaction, plus the elongated version SM2E of model SM2 in equations \eqref{eq:sabra_helical_general_up}-\eqref{eq:sabra_helical_general_um}. These coefficients ensure Energy and Helicity conservation. Conventionally, and without loss of generality, we always choose $a=1$.}
\label{tab:model_coefficients} 
\begin{tabular*}{\linewidth}{@{\extracolsep{\fill} } c  c  c }
    \toprule
    Model & b & c \\ \colrule
    SM1&$-1/2$&$1/2$\\
    SM2&$-5/2$&$-3/2$\\
		SM2E&$-9/4$&$-5/4$\\
    SM3&$-5/6$&$1/6$\\
    SM4&$-3/2$&$-1/2$\\
    \botrule
\end{tabular*}
\end{table}

\subsection{The \textit{elongated} helical SABRA model}
We introduce here a shell-model, which we will call SM2E, in which the triads are more elongated, 
in the sense that the middle wave-number $k_n$ in a triad will interact with the first 
larger neighbor $k_{n+1}$ and the second smaller neighbor $k_{n-2}$. In this way we have the equivalent of the parameter $v=k_{n-2}/k_n=\lambda^{-2}=0.25$ instead of $v=0.5$ as for the local version. The model equations take the form:
\begin{align}
\label{eq:sabra_helical_general_up}
\dot{u}_n^+ = & i (a k_{n+2} u_{n+3}^{s_1} u_{n+2}^{s_2*} + b k_{n}u_{n+1}^{s_3} u_{n-2}^{s_4*} + c k_{n-1} u_{n-1}^{s_5} u_{n-3}^{s_6}) - \nu k_n^\beta  u_n^+ + f_n^+ 
-\nu_l k_n^{-4} u_n^+ \, ,\\
\label{eq:sabra_helical_general_um}
\dot{u}_n^- = & i (a k_{n+2} u_{n+3}^{-s_1} u_{n+2}^{-s_2*} + b k_{n} u_{n+1}^{-s_3} u_{n-2}^{-s_4*} + c k_{n-1} u_{n-1}^{-s_5} u_{n-3}^{-s_6}) -  \nu k_n^\beta u_n^- + f_n^- -\nu_l k_n^{-4} u_n^- \, ,
\end{align}
where the helical indices $s_i$ fall in  the same helical class of the SM2 model (see Table \ref{tab:helicity_indices}). The real constants $a,b,c$ are determined by imposing that the triadic interaction conserves energy \eqref{eq:energy_def} and helicity \eqref{eq:helicity_def}. The values of the resulting coefficients for the SM2E model are given in Table \ref{tab:model_coefficients}.

In Appendix \ref{sec:app_general_sabra_equations} we give the  equations for an even more general shell-model, allowing for interacting triads of any shape.

In the next section we extend  the ``instability assumption'' developed in \cite{waleffe_1992_The_nature_of_triad_interactions} to predict the transfer properties of helical shell models, and we show that, indeed, the elongated version SM2E of the model SM2 should lead to an inverse energy transfer in agreement with the predictions for the set of triads with a similar geometrical factor in the Navier-Stokes equations.

\section{Energy transfers in helical shell models}
\label{sec:energy_transfers}

In this section we will first study the stability of steady states of only one triad of wave-numbers. We will then extend the results of this analysis to a shell-model with any number $N$ of shells, in the framework of the instability assumption \cite{waleffe_1992_The_nature_of_triad_interactions}.
The instability assumption states two things: (i) the global statistical behavior of a shell-model can be inferred directly from the single-triad dynamics; (ii)  in  a single-triad system the energy flows from the most unstable wave-number to the other two. Here the adjective ``unstable'' is intended to be used in the framework of the linear stability analysis of the equations for  $u_n^\pm$. In fact, proceeding as in \cite{waleffe_1992_The_nature_of_triad_interactions} and \cite{benzi_1996_Helical_shell_models}, we studied the linear stability of a single-triad helical shell-model, both in its first-neighbor and elongated variants. This analysis (see Appendix \ref{sec:app_instability_assumption}) confirms that there is one unstable wave-number that transfers energy to the other two. For models SM1 and SM3 the unstable wave-number is the smallest one, while for models SM2, SM2E and SM4 the unstable wave-number is the middle one (this property depends only on the helical class of the model, not on the triad shape). These results, already discussed in \cite{benzi_1996_Helical_shell_models}, are the same as those obtained for the Navier-Stokes equations, and they are summarized in Fig. \ref{fig:triad_stability}.

\subsection{Energy transfers}
Let us now examine how one can exploit the stability analysis for a single triad to 
predict the sign of the energy transfer in a fully coupled shell-model. 
For the balance of energy at shell $k_n$ we have:
\begin{equation}
\label{eq:energy_evolution}
\dot{E_n} = \frac{d}{dt} (|u_n^+|^2 + |u_n^-|^2 ) =
 ( \delta^E_{n+m} + b \delta^E_{n} - c \delta^E_{n-1})  - 2 \nu k_n^\beta E_n + 2 Re(f^+u_n^{+*} + f^-u_n^{-*} ) \, - 2 \nu_l k_n^{-4} E_n ,
\end{equation}
where 
\begin{equation}
\delta^E_n = -2 k_n Im[(u_{n+1}^{s_3} u_{n}^{+*} u_{n-m}^{s_4*}) + (u_{n+1}^{-s_3} u_{n}^{-*} u_{n-m}^{-s_4*})] \, ,
\end{equation}
and $m=1$ for the first-neighbor models SM1-SM4, or $m=2$ for model SM2E. 
The total energy flux across a shell $n$  is given by the balance equation: 
\begin{equation}
\sum_{j=0}^n\dot{E_j} = \Pi^E_n   - \epsilon^{out}_n + \epsilon^{in}_n  - \alpha^{out}_n  \, ,
\label{eq:flux}
\end{equation}
where the non-linear contribution is given by  $\Pi^E_n = \sum_{j=0}^n ( \delta^E_{j+m} + b \delta^E_{j} - c \delta^E_{j-1})$, and with 
$\epsilon^{out}_n = 2 \nu \sum_{j=0}^n k_j^\beta E_j $ and $\alpha^{out}_n = 2 \nu_l \sum_{j=0}^n k_j^{-4} E_j$ 
we denote the dissipative contributions at small and large scales respectively, 
 while with 
$\epsilon^{in}_n = 2 \sum_{j=0}^n   Re(f^+u_j^{+*} + f^-u_j^{-*} )  $ we denote the external input from the forcing. 
 Using the constraint of energy conservation $c=1+b$  (see  Appendix \ref{sec:app_general_sabra_equations}), one finds that the non-linear contribution to the flux can be further simplified for models SM1-SM4 to: 
\begin{equation}
\label{eq:e_flux_models_p1}
\Pi^E_n = (1 + b) \delta_{n}^{E} + \delta_{n+1}^{E}  \, ,
\end{equation}
while for model SM2E:
\begin{equation}
\label{eq:e_flux_m2p2}
\Pi^E_n =  (1 + b) \delta_{n}^{E} + \delta_{n+1}^{E} + \delta_{n+2}^{E} \, .
\end{equation}
The fact that the energy is conserved by the non-linear terms implies that the non-linear flux must vanish if calculated over all shells, $\Pi_N^E=0$. In the presence of a stationary statistics, an average of the left-hand side of \eqref{eq:flux} must vanish too. For the case of a  direct energy cascade ($\alpha_N^{out}\sim 0$), the global energy balance imposes the equality of the time-averaged values $\langle \epsilon^{out}_N \rangle = \langle \epsilon^{in}_N \rangle$, while for the  inverse energy cascade ($\epsilon^{out}_N \sim 0$)
  we must have $\langle \alpha^{out}_N \rangle = \langle \epsilon^{in}_N \rangle$. In the presence of a direct cascade and in 
the inertial range of scales, i.e. for wavenumbers, $k_n$,  much larger than the forcing scales, $k_{f}$, and much smaller than the viscous  scale, $k_\eta$, we must also have  $\epsilon_n^{out}= \alpha_n^{out} \sim 0$ and $\langle \epsilon_n^{in} \rangle = \text{const}$. As a consequence, the existence of a constant direct  energy cascade implies that  
$\langle \delta_{n}^{E} \rangle$ must be asymptotically  constant (independent of $n$), such that  also the flux will be constant and given by:
\begin{equation}
\label{eq:e_flux_raccolto}
	\langle \Pi^E_n \rangle = f(b) \langle \delta^{E}_n \rangle = - \langle \epsilon^{in}_n \rangle= \text{const} \, ,
\end{equation}
where $	f(b) = (2 + b) $ for models SM1-SM4, and $f(b) = (3 + b)$ for model SM2E. Similarly, in the presence of an inverse energy cascade regime, for wavenumbers $k_n$ smaller than $k_f$ we must have 
\begin{equation}
\label{eq:e_flux_raccolto1}
	\langle \Pi^E_n \rangle = f(b) \langle \delta^{E}_n \rangle = \langle \alpha^{out}_n \rangle = \text{const} \, .
\end{equation}
 
In  our notation, a negative flux means that  energy is flowing from large to small scales and vice versa. The sign of $f(b)$ is known once a model is chosen, while for finding the sign of $\langle \delta_n^{E} \rangle$ we make use of the instability assumption as follows.

Given a one-triad model, considering only shells $k_{n-m},k_n,k_{n+1}$, with zero energy injection and dissipation, equations \eqref{eq:energy_evolution}, after averaging and under the hypothesis of constant flux, will take the form:
\begin{align}
\label{eq:energy_variation_triad_with_const_flux}
\langle \dot{E}_{n-m} \rangle & = \langle \delta^E_n \rangle \, ,\nonumber \\
\langle \dot{E}_n \rangle & = b \langle \delta^E_n \rangle \, ,\nonumber \\
\langle \dot{E}_{n+1} \rangle & = - c \langle \delta^E_n \rangle \, . 
\end{align}
We can make several considerations based on equations \eqref{eq:energy_variation_triad_with_const_flux}. First, the ratio between the energies flowing towards the two stable wave-numbers is simply given by the $b$ and $c$ coefficients of the model. Second, exploiting the instability assumption (Fig. \ref{fig:triad_stability}), we can predict which wave-number should have positive or negative energy variation (the unstable will have a negative energy derivative and vice versa); since $b$ and $c$ are known (Table \ref{tab:model_coefficients}), the sign of $\langle \delta^{E}_n \rangle$ can be readily calculated. For instance, for model SM1, the mode with the smallest  wave-number is unstable, providing $\langle \dot{E}_{n-1} \rangle < 0$, $\langle \dot{E}_{n} \rangle > 0$ and $\langle \dot{E}_{n+1} \rangle > 0$; the values $b=-1/2<0$ and $c=1/2>0$ in equation \eqref{eq:energy_variation_triad_with_const_flux} yield $\langle \delta^{E}_n \rangle < 0$. Similarly, for model SM3 we have $\langle \delta^{E}_n \rangle < 0$, while for models SM2 and SM4 $\langle \delta^{E}_n \rangle > 0$. These results do not depend on the triad shape, but only on the helical class of the interaction, so also for model SM2E $\langle \delta^{E}_n \rangle > 0$. From these calculations, and equation \eqref{eq:e_flux_raccolto} we derive the predictions for the direction of the energy flux given in Table \ref{tab:flux_prediction}.

\begin{table}
\caption{Predictions for the energy flux, based on the instability assumption and equations \eqref{eq:flux}, \eqref{eq:e_flux_raccolto} and \eqref{eq:e_flux_raccolto1}. A negative energy flux means that energy is cascading towards small scales and vice versa. $\text{sgn}[x]$ is the sign function.}
\label{tab:flux_prediction} 
\begin{tabular*}{\linewidth}{@{\extracolsep{\fill} } c  c  c  c}
    \toprule
    Model & $\text{sgn}[\langle \delta^{E}_n \rangle]$ & $\text{sgn}[f(b)]$ & Energy flux prediction \\ \colrule
   SM1  & $-$ & $+$ &  Forward  \\
   SM2  & $+$ & $-$ &  Forward  \\
	 SM2E & $+$ & $+$ &  Backward \\
   SM3  & $-$ & $+$ &  Forward  \\
   SM4  & $+$ & $+$ &  Backward \\
    \botrule
\end{tabular*}
\end{table}

In this formalism, the information regarding the shape of the triad, i.e., the degree of non-locality, is entirely contained in the $f(b)$ prefactor. In order to have a positive energy flux in Eq. \eqref{eq:e_flux_raccolto1}, corresponding to an inverse energy cascade, the signs of the factors $\langle \delta^{E}_n \rangle$ and $f(b)$ must be the same.
We see that the above argument predicts that model SM4 will have a positive energy flux and would be the first candidate for a shell-model that displays inverse energy cascade. 
As shown in \cite{gilbert2002inverse}, it turns out that the fluctuations of the energy flux are so strong that such a system shows quasi-equilibrium rather than an inverse cascade of energy. 
However, also switching from model SM2 to SM2E, the energy flux should reverse its sign, due to the sign change in the factor $f(b)$, as predicted also for the NS case. This provides a good candidate for a model with the same invariants as 3D Navier-Stokes equations exhibiting inverse energy cascade.

\section{Numerical Simulations}
\label{sec:simulations_single}
In order to test the predictions made in section \ref{sec:energy_transfers}, and especially to see if the transition from the local shell model SM2 to the elongated shell model SM2E actually shows a reversal in the direction of the energy cascade, we have performed several numerical integrations of the equations \eqref{eq:sabra_helical_standard_up}-\eqref{eq:sabra_helical_standard_um} and \eqref{eq:sabra_helical_general_up}-\eqref{eq:sabra_helical_general_um}.
The energy is injected through a stochastic Gaussian forcing, delta correlated in time, with zero mean and $O(1)$ standard deviation, on two shells, both on the positive ($u_n^+$) and negative ($u_n^-$) helicity-carrying velocities, with different amplitudes, in order to inject helicity as well. 
We performed several simulations, with energy injected at large, medium or small scales, and for some of these cases we used hyper-viscosity ($\sim k^4$ dissipative term) in order to have a cleaner inertial range without increasing too much the number of shells. We wanted to verify that this hyper-viscosity does not have any important effect on the scaling laws of the observables. Also, a large-scale energy dissipation of the form $\sim k^{-4}$ was introduced in order to avoid large-scale energy accumulation where necessary. The parameters used for the simulations can be found in 
Table  \ref{tab:parameters_all}.
\begin {table*}
\caption{Parameters used for the simulations. Several simulations were performed with energy injected at different shell-numbers $k_{n_f}$. 
$|f_n^+|$ and $|f_n^+|$ represent the intensity (standard deviation) of the Gaussian forcing on the positive and negative helicity-carrying shells, respectively. Large scale dissipation: $\nu_l k_n^{-4}$. Small scale dissipation: sets I and II use a standard $\nu k_n^2$ viscosity, while sets III and IV use a $\nu k_n^4$ hyper-viscosity. For all 
 runs $\lambda=2$ and $k_0=1$.}
\label{tab:parameters_all} 
\begin{tabular*}{\linewidth}{@{\extracolsep{\fill} } c  c  c  c  c  c  c  c  }
    \toprule
      & $N$ & $\Delta t$ & $\nu$ & $\nu_l$ & $n_f$ & $|f_n^+|$ & $|f_n^-|$  \\ \colrule
    Run I & $36$ & $5 \cdot 10^{-9}$ & $1.0 \cdot 10^{-12}$ & $1$ & $4,5$ & $1$ & $0.5$  \\
    Run II & $36$ & $1 \cdot 10^{-8}$ & $1.0 \cdot 10^{-12}$ & $1$ & $4,5$ & $1$ & $0.5$  \\
    Run III & $31$ & $5 \cdot 10^{-9}$ & $2.5 \cdot 10^{-28}$ & $1$ & $14,15$ & $1$ & $0.5$  \\
    Run IV & $31$ & $1 \cdot 10^{-8}$ & $2.5 \cdot 10^{-28}$ & $1$ & $22,23$ & $1$ & $0.5$  \\
    \botrule
\end{tabular*}
\end{table*}

The time integration has been carried out using an explicit 2nd-order Adams-Basforth scheme with exact integration of the viscous terms:
\begin{equation}
\label{eq:integration_scheme}
	u_n(t+\Delta t) = u_n(t) e^{-\gamma_n \Delta t} + \Delta t \left[ \frac{3}{2} e^{-\gamma_n \Delta t} NLT_n(t) -\frac{1}{2} e^{-2\gamma_n \Delta t} NLT_n(t-\Delta t) \right] \, ,
\end{equation}
where $\gamma_n$ and $NLT_n$ are, respectively, the viscous and the non-linear terms on the right hand side of \eqref{eq:sabra_helical_standard_up}-\eqref{eq:sabra_helical_standard_um} or \eqref{eq:sabra_helical_general_up}-\eqref{eq:sabra_helical_general_um}. 
The stochastic forcing is integrated separately with a forward Euler scheme.

\begin{figure}
\centering
\includegraphics{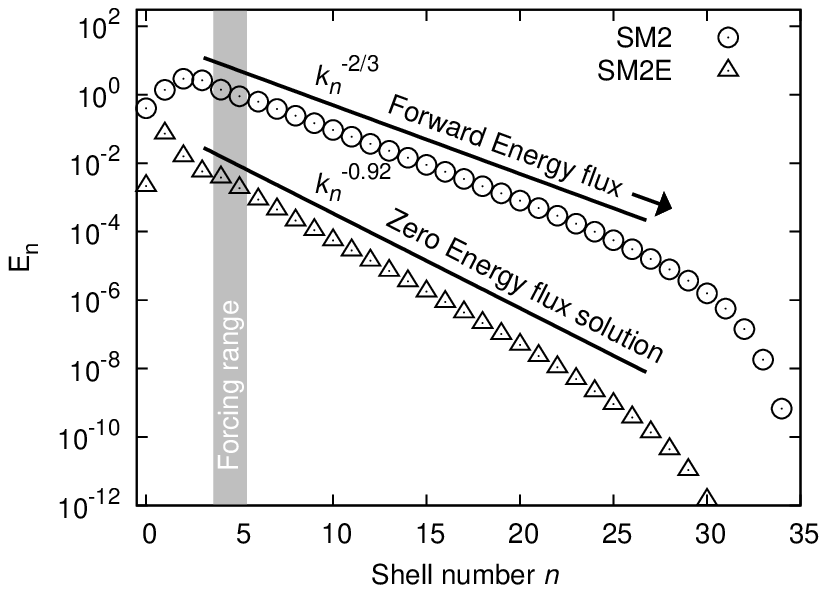}
\caption{Energy spectra $E_n$ for the two variants of model 2, forced at large scales (gray shaded region). Curves are shifted vertically for clarity. Parameters used for this simulation are in Table \ref{tab:parameters_all} (runs I and II).} 
\label{fig:multiple_e_spectra_large} 
\end{figure}

\begin{figure}
\centering
\includegraphics{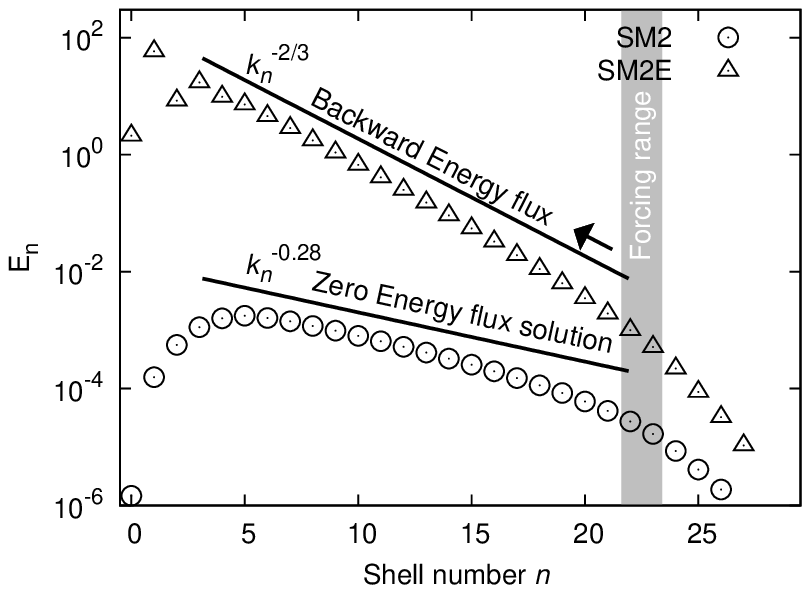}
\caption{Energy spectra $E_n$ for the two variants of model 2, forced at small scales (gray shaded region). Curves are shifted vertically for clarity. Parameters used for this simulation are in Table \ref{tab:parameters_all}  (run IV).} 
\label{fig:multiple_e_spectra_small} 
\end{figure}

\begin{figure}
\centering
\includegraphics{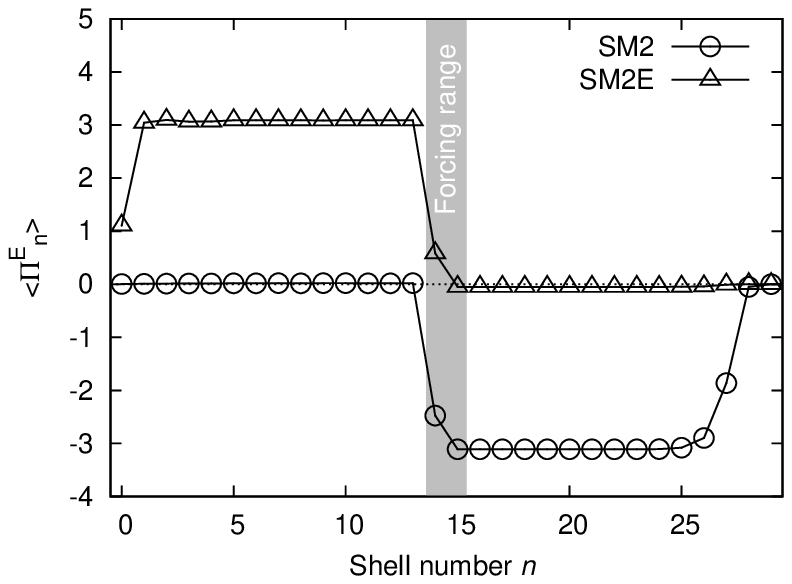}
\caption{Energy flux $\langle \Pi^E_n \rangle$ for the two variants of model 2, forced at medium scales (gray shaded region). We recall that, with our notation, a positive energy flux corresponds to an inverse cascade of energy, and vice-versa. Parameters used for this simulation are in Table \ref{tab:parameters_all} (run III).}
\label{fig:multiple_e_flux_medium} 
\end{figure}

The equations were evolved for several hundreds of large-scale  eddy turnover times, $T_e$,
and time averages have been first calculated on runs lasting $T \sim 10 T_e$ and then averaged over all the stationary runs. Stationarity is checked by monitoring the total energy evolution.
Figures \ref{fig:multiple_e_spectra_large} and \ref{fig:multiple_e_spectra_small} show the energy spectra for the local SM2 and elongated SM2E models, for both large-scale and small-scale energy injection cases.
Figure \ref{fig:multiple_e_flux_medium}  shows the corresponding energy flux for the case when the forcing mechanism is acting at an intermediate scale, such as to resolve simultaneously the 
forward and backward transfers. 
We briefly remind the reader that in terms of shell-model variables, a forward/backward energy cascade 
gives the scaling $E_n \sim |\epsilon|^{2/3} k^{-2/3}$, while a dynamics close to energy equipartition
should give  $E_n \sim \text{const}$.

From Figs. \ref{fig:multiple_e_spectra_large} and  \ref{fig:multiple_e_flux_medium}  we clearly see that model SM2 has a forward energy transfer and no backward transfer. On the other hand, 
Figs. \ref{fig:multiple_e_spectra_small} and  \ref{fig:multiple_e_flux_medium}
 show that model SM2E has the opposite behavior: a clear backward energy transfer and zero forward flux.  
Let us further notice that in the regime where the energy flux is absent both models do not develop a solution close to energy equipartition. 
Indeed,  in these ranges the dynamics can be dominated by a homogeneous solution of the energy balance equation \eqref{eq:energy_evolution} in the stationary regime.
 In fact, by substituting the definitions \eqref{eq:e_flux_models_p1} or \eqref{eq:e_flux_m2p2} inside 
the stationary balance equation for the flux, $\langle \Pi^E_n \rangle = \epsilon$, where the sign of $\epsilon$ depends on whether we have a forward or a backward cascade,  we obtain the following for the two models SM2 and SM2E:

\begin{align}
\label{eq:e_flux_models_p1_zeromodes}
	(1 + b) \langle \delta_{n}^{E} \rangle + \langle \delta_{n+1}^{E} \rangle &= \epsilon \, , \qquad \text{(SM2)} \\
\label{eq:e_flux_m2p2_zeromodes}
	(1 + b) \langle \delta_{n}^{E} \rangle + \langle \delta_{n+1}^{E} \rangle + \langle \delta_{n+2}^{E} \rangle &= \epsilon \, . \qquad \text{(SM2E)}
\end{align}

The solution of these difference equations is generally a sum of the solution of the corresponding homogeneous equation (zero-flux solution, or \textit{zero-mode}) and a particular solution, for instance $\langle \delta_n^E  \rangle =\text{const}$, that represents the constant flux solution \cite{bender_orszag_book}. If the homogeneous solutions has a steeper scaling than the constant-flux solutions, it is  sub-dominant in the dynamics. On the other hand, when the constant energy flux solution is absent, the homogeneous zero-mode may become dominant.
 This explains the slope of the energy spectrum for the SM2 model in the range $k<k_f$, where a direct calculation shows that 
the dynamics is dominated by a zero-mode solution of \eqref{eq:e_flux_models_p1_zeromodes}, $\langle \delta_{n+1}^{E} \rangle/\langle \delta_{n}^{E} \rangle = -(1+b) = \lambda^{0.585}$, leading to the scaling law $|u_n|^2 \sim (\langle \delta_{n}^{E} \rangle/k_n)^{2/3} \sim k_n^{0.277}$, see Fig. \ref{fig:multiple_e_spectra_small}. The same may also happen with the SM2E model, in the range $k>k_f$, where the  scaling imposed by the zero-mode, $|u_n|^2 \sim k_n^{-0.92}$, is very close to that observed in Fig. \ref{fig:multiple_e_spectra_large}. 

For completeness, we must say that there are situations in which the scaling dictated by the zero-mode of the energy flux is the same as the scaling given by the constant helicity flux solution. Also, the zero mode of the helicity flux may dictate the same scaling as that given by the constant energy flux solution. This happens for models SM1 and SM4.

\begin{figure}
\centering
\includegraphics{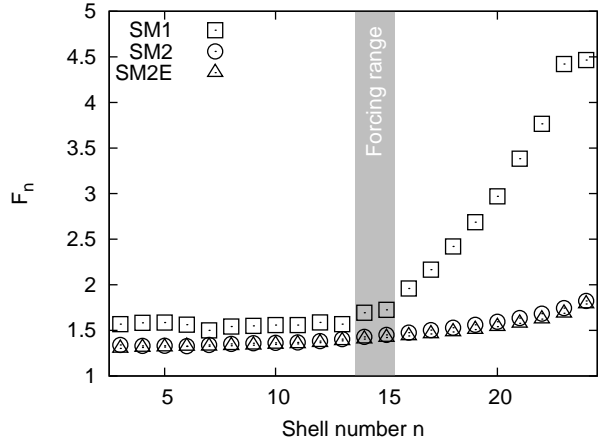}
\caption{Flatness $F_n$ for models SM1, SM2 and SM2E, forced at medium scales (gray shaded region). Model SM1 is intermittent in the $n>n_f$ range. Models SM2 and SM2E show very weak level of intermittency for $n>n_f$. Parameters used for this simulation are in Table \ref{tab:parameters_all} (run III).}
\label{fig:flatness_medium} 
\end{figure}

Another interesting question is about intermittency. It is generally believed that inverse cascades
 do not show any anomalous scaling, i.e. they are not intermittent, while forward cascades do.  
One way of quantifying intermittency is by looking at the flatness, the ratio between the fourth-order moment and the squared second-order moment, as a function of the reference scale. Figure \ref{fig:flatness_medium} shows the flatness of the total shell energy  defined as
\begin{equation}
	F_n = \frac{S_4(k_n)}{[S_2(k_n)]^{2}}
\end{equation}
for models SM1, SM2, and SM2E, where the structure functions $S_q(k_n)$ are defined in terms of the energy flux \eqref{eq:e_flux_models_p1} and \eqref{eq:e_flux_m2p2}:
\begin{equation}
	S_q(k_n) = \left\langle (k_n^{-1}|\Pi^E_n|)^{\frac{q}{3}} \right\rangle \, .
\end{equation}
The larger the values of the flatness, the more non-Gaussian is the PDF. As one can see, model SM1 develops a clear anomalous scaling in the forward regime (for $k_n > k_f$) and no intermittency for $k<k_f$, where it is known to be dominated by equilibrium statistics (no backward energy transfer). Note that model SM1 can be shown to be equivalent to the original SABRA model, which is known to have an intermittent dynamics in the $n>n_f$ range. Models SM2 and SM2E have very little visible deviations in the forward regimes and no intermittency at all in the backward regime in agreement with the observation that inverse cascades do not develop anomalous scaling \cite{paret1998intermittency, boffetta2000inverse}. These results are generally interpreted in term of the hierarchy of time scales in the system: a forward energy cascade with spectrum $E_n \sim k_n^{-2/3}$ implies that the typical eddy turn-over time at shell $n$ goes like $\tau_n \sim 1/(k_n u_n) \sim k_n^{-2/3}$, i.e. energy is transferred to 
faster and faster modes, preventing small-scales to equilibrate around the mean properties of the large ones. On the other hand, an inverse energy cascade with the same slope is dominated by exactly the opposite dynamics, i.e., fast scales transfer fluctuations to slower ones allowing for self-averaging. It is not clear if this phenomenology is at the root also of shell models dynamics, where energy is known to be transferred also via quasi-instantonic solutions traveling coherently among a huge set of shell variables \cite{dombre1998intermittency, biferale1999multi, daumont2000instanton, constantin2007regularity, mailybaev2013blowup}. This argument is the aim of a work in progress, and it will be reported elsewhere. 

Finally,  for models having helical interaction SM2, in order to check that the reversal in the energy flux is robust when the ratio of the smallest to the middle wave number is $v<0.278$ \cite{waleffe_1992_The_nature_of_triad_interactions}, we simulated numerically another model, with first-neighbor and third-neighbor interactions ($k_{n-3}, k_n, k_{n+1}$) (see Appendix \ref{sec:app_general_sabra_equations}). For this model, $v=0.125$, and the results for the energy spectrum, the energy fluxes, and the intermittency are qualitatively the same as for the model SM2E (not shown).

\section{Conclusions}
We have generalized a previously proposed class of helical shell models to include also non-local triadic interactions among Fourier modes. Using arguments similar to those developed for the Navier-Stokes equations \cite{waleffe_1992_The_nature_of_triad_interactions} we have shown that a suitable subset of helical triadic interactions may change the energy transfer direction depending on the relative geometry of the three interacting modes, leading to direct or inverse cascades.
We also show that the inverse cascade is not intermittent and that the scaling properties in the range of shells where the energy does not flow might be dominated by a zero-mode solution 
of the energy balance equations.
This work opens the way to study the coupling between different models with different helical interactions and triad shapes in order to understand and mimic those transitions from direct to inverse cascades observed in real flows at changing the degree of rotation, aspect ratio or large scale shear \cite{staplehurst2008structure, kellay2002two}. Coupling models with different transfer properties makes it more challenging to disentangle the effects of the dynamics coming from each single model. Schemes such as mode-to-mode energy transfer \cite{dar2001energy, verma2004statistical} can be efficiently combined with our formalism to address this issue.
Another interesting direction for future work is to understand the influence of high helicity content on the dynamics of direct and inverse triadic interactions, or, in general, the dynamics of the helicity in the inverse cascade model, as was done in \cite{lessinnes2011dissipation} for the direct cascade model SM3.

\section*{Acknowledgments}
M.D.P. and L.B. acknowledge funding from the European Research Council under the European Union's Seventh Framework Programme, ERC AdG NewTURB Agreement No 339032. A.A.M. was supported by the CNPq (grant 305519/2012-3) and by the FAPERJ (Pensa Rio 2014).

\appendix

\section{Equations and coefficients for a helical SABRA shell-model with generic wave-number triads}
\label{sec:app_general_sabra_equations}

In this Appendix we present the equations for a SABRA shell-model with a generic triad shape. For the sake of simplicity we omit the forcing and the dissipative terms. 
 The equations are:
\begin{align}
\label{eq:sabra_most_general}
\frac{d}{dt}u_n^+ &= i ( a k_{n+m} u_{n+m+l}^{s_1} u_{n+m}^{s_2*} + b k_{n} u_{n+l}^{s_3} u_{n-m}^{s_4*} + c k_{n-l} u_{n-l}^{s_5} u_{n-m-l}^{s_6})  \, , \nonumber \\
 \frac{d}{dt} u_n^- &= i ( a k_{n+m} u_{n+m+l}^{-s_1} u_{n+m}^{-s_2*} + b k_{n} u_{n+l}^{-s_3} u_{n-m}^{-s_4*} + c k_{n-l} u_{n-l}^{-s_5} u_{n-m-l}^{-s_6})  \, .
\end{align}
Here $a,b,c$  are real coefficients, the helical indices $s_i = \pm$ are reported in Table \ref{tab:helicity_indices} and the triad shape ($k_{n-m}, k_n, k_{n+l}$) is described by the pair of indices $m$ and $l$.
The coefficient $a$ can always be set equal to $1$ just by rescaling the other coefficients and time. The coefficients $b$ and $c$ are fixed by imposing the conservation of the quadratic inviscid  invariants as follows.

It can be shown that equations \eqref{eq:sabra_most_general} admit only four quadratic inviscid invariants. Only two out of the four can be simultaneously conserved, due to the fact that there are only two 
free parameters ($b$ and $c$). The four possible invariants are:
\begin{enumerate}
	\item $W^I \equiv \sum_n k_n^{\alpha_I}(|u^+|^2 + |u^-|^2)$, which for $\alpha_I = 0$ corresponds to the total energy.
	
	\item $W^{II} \equiv \sum_n k_n^{\alpha_{II}}(|u^+|^2 - |u^-|^2)$, which is not sign-definite and for $\alpha_{II} = 1$
 corresponds to the total helicity.
	
	\item $W^{III} \equiv \sum_n (-1)^n k_n^{\alpha_{III}}(|u^+|^2 + |u^-|^2)$.

	\item $W^{IV} \equiv \sum_n (-1)^n k_n^{\alpha_{IV}}(|u^+|^2 - |u^-|^2)$.
	
\end{enumerate}
Only  $W^I$ and $W^{II}$ have the same physical meaning as the invariants of the NS equations, while for $W^{III}$ and $W^{IV}$ there is no such analogy.

The triad-by-triad conservation of a $W^{I}$-type invariant implies:

\begin{equation}
0=\frac{d}{dt} \sum_n k_n^\alpha (|u_n^+|^2 + |u_n^-|^2 ) = \sum_n k_n^\alpha ( \dot{u_n}^+ u_n^{+*} + \dot{u_n}^- u_n^{-*} + c.c.) \, ,
\end{equation}
where all the terms on the right-hand side must formally cancel for each triad after substituting equations \eqref{eq:sabra_most_general}. For all the 4 classes of helical interaction the resulting conservation equation is:

\begin{equation}
\label{eq:energy_conservation_coefficients}
a + \lambda^{\alpha_I m} b - \lambda^{\alpha_I (m+l)} c = 0 \, .
\end{equation}

The conservation of a $W^{II}$-type invariant yields respectively:

\begin{eqnarray}
\label{eq:helicity_conservation_coefficients_m1}
a - \lambda^{\alpha_{II} m} b - \lambda^{\alpha_{II} (m+l)} c = 0 \, , \quad \text{(SM1)} \\
\label{eq:helicity_conservation_coefficients_m2}
a - \lambda^{\alpha_{II} m} b + \lambda^{\alpha_{II} (m+l)} c = 0 \, , \quad \text{(SM2)} \\
\label{eq:helicity_conservation_coefficients_m3}
a + \lambda^{\alpha_{II} m} b + \lambda^{\alpha_{II} (m+l)} c = 0 \, , \quad \text{(SM3)} \\
\label{eq:helicity_conservation_coefficients_m4}
a + \lambda^{\alpha_{II} m} b - \lambda^{\alpha_{II} (m+l)} c = 0 \, . \quad \text{(SM4)} 
\end{eqnarray}

As said before, we can always choose $a=1$. Solving equations \eqref{eq:energy_conservation_coefficients} and \eqref{eq:helicity_conservation_coefficients_m1}-\eqref{eq:helicity_conservation_coefficients_m4} for each model, we get the generic expressions for the $b$ and $c$ coefficients, which are reported in Table \ref{tab:model_coefficients_generic}.

\begin {table}[Hb]
\caption{General expression for the coefficients of equations \eqref{eq:sabra_most_general} conserving generic invariants $W^{I}$ and $W^{II}$. Without loss of generality $a=1$. For models that conserve Energy and Helicity, one should set $\alpha_I = 0$ and $\alpha_{II} = 1$.}
\label{tab:model_coefficients_generic} 
\begin{tabular*}{\linewidth}{@{\extracolsep{\fill} } c  c  c }
    \toprule
    Model & b & c \\ \colrule
    1 & $ \frac{ \lambda^{\alpha_I(m+l)} [1-\lambda^{(m+l)(\alpha_{II}-\alpha_I)}] }{ \lambda^{m(\alpha_I+\alpha_{II})} (\lambda^{l \alpha_{II}} + \lambda^{l \alpha_{I}}) }$ & $- \frac{ \lambda^{m \alpha_I} [-1-\lambda^{m(\alpha_{II}-\alpha_I)}] }{ \lambda^{m(\alpha_I+\alpha_{II})} (\lambda^{l \alpha_{II}} + \lambda^{l \alpha_{I}}) }$\\
    2 & $ \frac{ \lambda^{\alpha_I(m+l)} [-1-\lambda^{(m+l)(\alpha_{II}-\alpha_I)}] }{ \lambda^{m(\alpha_I+\alpha_{II})} (\lambda^{l \alpha_{II}} - \lambda^{l \alpha_{I}}) }$ & $ \frac{ \lambda^{m \alpha_I} [-1-\lambda^{m(\alpha_{II}-\alpha_I)}] }{ \lambda^{m(\alpha_I+\alpha_{II})} (\lambda^{l \alpha_{II}} - \lambda^{l \alpha_{I}}) }$\\
    3 & $ \frac{ \lambda^{\alpha_I(m+l)} [-1-\lambda^{(m+l)(\alpha_{II}-\alpha_I)}] }{ \lambda^{m(\alpha_I+\alpha_{II})} (\lambda^{l \alpha_{II}} + \lambda^{l \alpha_{I}}) }$ & $ \frac{ \lambda^{m \alpha_I} [-1+\lambda^{m(\alpha_{II}-\alpha_I)}] }{ \lambda^{m(\alpha_I+\alpha_{II})} (\lambda^{l \alpha_{II}} + \lambda^{l \alpha_{I}}) }$\\
    4 & $ \frac{ \lambda^{\alpha_I(m+l)} [-1+\lambda^{(m+l)(\alpha_{II}-\alpha_I)}] }{ \lambda^{m(\alpha_I+\alpha_{II})} (-\lambda^{l \alpha_{II}} + \lambda^{l \alpha_{I}}) }$ & $ \frac{ \lambda^{m \alpha_I} [-1+\lambda^{m(\alpha_{II}-\alpha_I)}] }{ \lambda^{m(\alpha_I+\alpha_{II})} (-\lambda^{l \alpha_{II}} + \lambda^{l \alpha_{I}}) }$\\
    \botrule
\end{tabular*}
\end{table}

\section{Instability assumption}
\label{sec:app_instability_assumption}

For completeness, we repeat here the calculations done in \cite{benzi_1996_Helical_shell_models} for the linear stability analysis of a triad of interacting wave-numbers.
Let us consider a system made of three consecutive wave-numbers $k_1$, $k_2=\lambda k_1$, and $k_3=\lambda^2 k_1$ ($\lambda>1$), and, for instance, model SM1. The equations of motion \eqref{eq:sabra_helical_standard_up}-\eqref{eq:sabra_helical_standard_um} for such a system reduce to:

\begin{align}
\label{eq:sabra_helical_one_triad}
\dot{u_1}^+ & = i k_2 a u_{3}^{+} u_{2}^{-*} \, ,\nonumber \\
\dot{u_2}^{-} & = i k_2 b  u_{3}^{+} u_{1}^{+*} \, ,\nonumber \\
\dot{u_3}^{+} & = i k_2 c u_{2}^{-} u_{1}^{+} \, .
\end{align}
This system has three equilibrium states, of the form $(u_1^+, u_2^-, u_3^+) \in \{(A,0,0) ;(0,A,0); (0,0,A)\}$, where $A \in \mathbb{C}$. Linearization of the system  around a generic state ($u_{n}^{s} \rightarrow u_{n}^{s} + \Delta_{n}^{s}$, with $\Delta_{n}^{s} \ll 1$) gives:

\begin{align}
\label{eq:sabra_helical_one_triad_perturbation}
\dot{\Delta_1}^+ & = i k_2 a (\Delta_{2}^{-*} u_{3}^{+} + \Delta_{3}^{+} u_{2}^{-*}) \, ,\nonumber \\
\dot{\Delta_2}^{-} & = i k_2 b  (\Delta_{1}^{+*} u_{3}^{+} + \Delta_{3}^{+} u_{1}^{+*}) \, ,\nonumber \\
\dot{\Delta_3}^{+} & = i k_2 c  (\Delta_{1}^{+} u_{2}^{-} + \Delta_{2}^{-} u_{1}^{+}) \, .\\
\end{align}
The eigenvalues relative to the first state $(A,0,0)$ are:
\begin{equation}
\label{eq:perturbation_1}
\lambda_1 = 0 \, , \quad \lambda_{2,3} = \pm k_2 |A| \sqrt{-bc} = \pm k_2 |A|/2 \, ,
\end{equation}
where we substituted $-bc=1/4$ (see table \ref{tab:model_coefficients}), hence the equilibrium state is unstable because one of the perturbations grows exponentially in time as $\Delta_i \sim \exp(k_2 |A| t /2)$.

Similarly, the eigenvalues relative to the second state $(0,A,0)$ are:
\begin{equation}
\label{eq:perturbation_2}
\lambda_1 = 0 \, , \quad \lambda_{2,3} = \pm k_2 |A| \sqrt{-ac} = \pm i k_2 |A|/\sqrt{2} \, ,
\end{equation}
so all the perturbations $\Delta_i$ are bounded in time. The same can be said for the third state $(0,0,A)$,  for which the eigenvalues are: 
\begin{equation}
\label{eq:perturbation_3}
\lambda_1 = 0 \, , \quad \lambda_{2,3} = \pm k_2 |A| \sqrt{ab} = \pm i k_2 |A|/\sqrt{2} \, .
\end{equation}
According to the terminology of \cite{waleffe_1992_The_nature_of_triad_interactions}, a wave-number $k_1$ represented by the unstable equilibrium state $(A,0,0)$, where the energy is flowing towards the other modes $k_2$ and $k_3$, is called unstable. Similarly, wave-numbers $k_2$ and $k_3$ are stable with respect to small perturbations, as suggested by \eqref{eq:perturbation_2} and \eqref{eq:perturbation_3}. So we see that for model SM1, the unstable wave-number is the smallest one. Furthermore, the stability depends only on the sign of the coefficients $a,b,c$, which again depends only on the type of helical interaction chosen, while the triad shape does not play any role. In fact, repeating the same calculations with a different triad shape gives exactly the same stability results.

Analogous equations can be written for the other models, and it is found that for model SM3 the unstable wave-number is the smallest one, while for models SM2, SM2E and SM4 the unstable wave-number is the middle one, as summarized in Fig. \ref{fig:triad_stability}.

\bibliographystyle{unsrt}
\bibliography{bibliography}

\begin{thebibliography}{10}

\bibitem{pope_turbulent_flows}
S.B. Pope.
\newblock {\em Turbulent Flows}.
\newblock Cambridge University Press, 2000.

\bibitem{frish_turbulence}
U.~Frisch.
\newblock {\em Turbulence: The Legacy of A. N. Kolmogorov}.
\newblock Cambridge University Press, 1995.

\bibitem{obukhov1971some}
A.~Obukhov.
\newblock Some general characteristic equations of the dynamics of the
  atmosphere.
\newblock {\em Izvestiya Akademii Nauk SSSR, Fizika Atmosfery i Okeana},
  7:695--704, 1971.

\bibitem{gledzer1973system}
E.B. Gledzer.
\newblock System of hydrodynamic type admitting two quadratic integrals of
  motion.
\newblock {\em Soviet Physics Doklady}, 18:216, 1973.

\bibitem{desnianskii1974evolution}
V.N. Desnianskii and E.A. Novikov.
\newblock Evolution of turbulence spectra toward a similarity regime.
\newblock {\em Akademiia Nauk SSSR, Izvestiia, Fizika Atmosfery i Okeana},
  10:127--136, 1974.

\bibitem{yamada1988inertial}
M.~Yamada and K.~Ohkitani.
\newblock {The inertial subrange and non-positive Lyapunov exponents in
  fully-developed turbulence}.
\newblock {\em Progress of Theoretical Physics}, 79(6):1265--1268, 1988.

\bibitem{jensen1991intermittency}
M.H. Jensen, G.~Paladin, and A.~Vulpiani.
\newblock Intermittency in a cascade model for three-dimensional turbulence.
\newblock {\em Physical Review A}, 43(2):798, 1991.

\bibitem{Lvov_1998_improved_shellmodels}
V.S. L'vov, E.~Podivilov, A.~Pomyalov, I.~Procaccia, and D.~Vandembroucq.
\newblock Improved shell model of turbulence.
\newblock {\em Phys. Rev. E}, 58:1811--1822, 1998.

\bibitem{biferale2003shell}
L.~Biferale.
\newblock Shell models of energy cascade in turbulence.
\newblock {\em Annual Review of Fluid Mechanics}, 35(1):441--468, 2003.

\bibitem{bohr2005dynamical}
T.~Bohr, M.H. Jensen, G.~Paladin, and A.~Vulpiani.
\newblock {\em Dynamical systems approach to turbulence}.
\newblock Cambridge University Press, 2005.

\bibitem{ditlevsen2010turbulence}
P.D. Ditlevsen.
\newblock {\em Turbulence and shell models}.
\newblock Cambridge University Press, 2010.

\bibitem{pisarenko1993further}
D.~Pisarenko, L.~Biferale, D.~Courvoisier, U.~Frisch, and M.~Vergassola.
\newblock Further results on multifractality in shell models.
\newblock {\em Physics of Fluids A: Fluid Dynamics (1989-1993)},
  5(10):2533--2538, 1993.

\bibitem{plunian2013shell}
F.~Plunian, R.~Stepanov, and P.~Frick.
\newblock Shell models of magnetohydrodynamic turbulence.
\newblock {\em Physics Reports}, 523(1):1--60, 2013.

\bibitem{reshetnyak2003shell}
M.~Reshetnyak and B.~Steffen.
\newblock The shell model approach to the rotating turbulence.
\newblock {\em arXiv preprint physics/0311001}, 2003.

\bibitem{hattori2004shell}
Y.~Hattori, R.~Rubinstein, and A.~Ishizawa.
\newblock Shell model for rotating turbulence.
\newblock {\em Physical Review E}, 70(4):046311, 2004.

\bibitem{chakraborty2010two}
S.~Chakraborty, M.H. Jensen, and A.~Sarkar.
\newblock On two-dimensionalization of three-dimensional turbulence in shell
  models.
\newblock {\em The European Physical Journal B}, 73(3):447--453, 2010.

\bibitem{brandenburg1992energy}
A.~Brandenburg.
\newblock Energy spectra in a model for convective turbulence.
\newblock {\em Physical Review Letters}, 69(4):605, 1992.

\bibitem{Mingshun1997Scaling}
J.~Mingshun and L.~Shida.
\newblock Scaling behavior of velocity and temperature in a shell model for
  thermal convective turbulence.
\newblock {\em Phys. Rev. E}, 56:441--446, Jul 1997.

\bibitem{Ching2008Anomalous}
Emily S.~C. Ching and W.~C. Cheng.
\newblock Anomalous scaling and refined similarity of an active scalar in a
  shell model of homogeneous turbulent convection.
\newblock {\em Phys. Rev. E}, 77:015303, Jan 2008.

\bibitem{Ching2008Refined}
Emily S.~C. Ching, H.~Guo, and T.~S. Lo.
\newblock Refined similarity hypotheses in shell models of homogeneous
  turbulence and turbulent convection.
\newblock {\em Phys. Rev. E}, 78:026303, Aug 2008.

\bibitem{Kumar2015Shell}
A.~Kumar and M.K. Verma.
\newblock {Shell model for buoyancy-driven turbulence}.
\newblock {\em Physical Review E}, 91(4):043014, 2015.

\bibitem{jensen1992shell}
M.H. Jensen, G.~Paladin, and A.~Vulpiani.
\newblock Shell model for turbulent advection of passive-scalar fields.
\newblock {\em Physical Review A}, 45(10):7214, 1992.

\bibitem{benzi1997analytic}
R.~Benzi, L.~Biferale, and A.~Wirth.
\newblock Analytic calculation of anomalous scaling in random shell models for
  a passive scalar.
\newblock {\em Physical Review Letters}, 78(26):4926, 1997.

\bibitem{arad2001statistical}
I.~Arad, L.~Biferale, A.~Celani, I.~Procaccia, and M.~Vergassola.
\newblock Statistical conservation laws in turbulent transport.
\newblock {\em Physical Review Letters}, 87(16):164502, 2001.

\bibitem{aurell1994statistical}
E.~Aurell, G.~Boffetta, A.~Crisanti, P.~Frick, G.~Paladin, and A.~Vulpiani.
\newblock Statistical mechanics of shell models for two-dimensional turbulence.
\newblock {\em Physical Review E}, 50(6):4705, 1994.

\bibitem{gilbert2002inverse}
T.~Gilbert, V.S. L’vov, A.~Pomyalov, and I.~Procaccia.
\newblock Inverse cascade regime in shell models of two-dimensional turbulence.
\newblock {\em Physical Review Letters}, 89(7):074501, 2002.

\bibitem{biferale2012inverse}
L.~Biferale, S.~Musacchio, and F.~Toschi.
\newblock Inverse energy cascade in three-dimensional isotropic turbulence.
\newblock {\em Physical Review Letters}, 108(16):164501, 2012.

\bibitem{benzi_1996_Helical_shell_models}
R.~Benzi, L.~Biferale, R.M. Kerr, and E.~Trovatore.
\newblock Helical shell models for three-dimensional turbulence.
\newblock {\em Phys. Rev. E}, 53:3541--3550, 1996.

\bibitem{waleffe_1992_The_nature_of_triad_interactions}
F.~Waleffe.
\newblock The nature of triad interactions in homogeneous turbulence.
\newblock {\em Physics of Fluids A}, 4(2):350--363, 1992.

\bibitem{ditlevsen2001cascades}
P.D. Ditlevsen and P.~Giuliani.
\newblock Cascades in helical turbulence.
\newblock {\em Physical Review E}, 63(3):036304, 2001.

\bibitem{chen2003joint}
Q.~Chen, S.~Chen, and G.L. Eyink.
\newblock The joint cascade of energy and helicity in three-dimensional
  turbulence.
\newblock {\em Physics of Fluids}, 15(2):361--374, 2003.

\bibitem{lessinnes2011dissipation}
T.~Lessinnes, F.~Plunian, R.~Stepanov, and D.~Carati.
\newblock Dissipation scales of kinetic helicities in turbulence.
\newblock {\em Physics of Fluids (1994-present)}, 23(3):035108, 2011.

\bibitem{bender_orszag_book}
C.M. Bender and S.A. Orszag.
\newblock {\em Advanced mathematical methods for scientists and engineers}.
\newblock International series in pure and applied mathematics. McGrew–Hill,
  New York, 1978.

\bibitem{paret1998intermittency}
J.~Paret and P.~Tabeling.
\newblock Intermittency in the two-dimensional inverse cascade of energy:
  Experimental observations.
\newblock {\em Physics of Fluids}, 10(12):3126--3136, 1998.

\bibitem{boffetta2000inverse}
G.~Boffetta, A.~Celani, and M.~Vergassola.
\newblock Inverse energy cascade in two-dimensional turbulence: Deviations from
  gaussian behavior.
\newblock {\em Physical Review E}, 61(1):R29, 2000.

\bibitem{dombre1998intermittency}
T.~Dombre and J.L. Gilson.
\newblock {Intermittency, chaos and singular fluctuations in the mixed
  Obukhov-Novikov shell model of turbulence}.
\newblock {\em Physica D: Nonlinear Phenomena}, 111(1):265--287, 1998.

\bibitem{biferale1999multi}
L.~Biferale, G.~Boffetta, A.~Celani, and F.~Toschi.
\newblock Multi-time, multi-scale correlation functions in turbulence and in
  turbulent models.
\newblock {\em Physica D: Nonlinear Phenomena}, 127(3):187--197, 1999.

\bibitem{daumont2000instanton}
I.~Daumont, T.~Dombre, and J.L. Gilson.
\newblock Instanton calculus in shell models of turbulence.
\newblock {\em Physical Review E}, 62(3):3592, 2000.

\bibitem{constantin2007regularity}
P.~Constantin, B.~Levant, and E.S. Titi.
\newblock Regularity of inviscid shell models of turbulence.
\newblock {\em Physical Review E}, 75(1):016304, 2007.

\bibitem{mailybaev2013blowup}
A.A. Mailybaev.
\newblock Blowup as a driving mechanism of turbulence in shell models.
\newblock {\em Physical Review E}, 87(5):053011, 2013.

\bibitem{staplehurst2008structure}
P.J. Staplehurst, P.A. Davidson, and S.B. Dalziel.
\newblock Structure formation in homogeneous freely decaying rotating
  turbulence.
\newblock {\em Journal of Fluid Mechanics}, 598:81--105, 2008.

\bibitem{kellay2002two}
H.~Kellay and W.I. Goldburg.
\newblock Two-dimensional turbulence: a review of some recent experiments.
\newblock {\em Reports on Progress in Physics}, 65(5):845, 2002.

\bibitem{dar2001energy}
Gaurav Dar, Mahendra~K Verma, and V~Eswaran.
\newblock Energy transfer in two-dimensional magnetohydrodynamic turbulence:
  formalism and numerical results.
\newblock {\em Physica D: Nonlinear Phenomena}, 157(3):207--225, 2001.

\bibitem{verma2004statistical}
Mahendra~K Verma.
\newblock Statistical theory of magnetohydrodynamic turbulence: recent results.
\newblock {\em Physics Reports}, 401(5):229--380, 2004.

\end{thebibliography}

\end{document}